# Detecting gene innovations for phenotypic diversity across multiple genomes

Inti Pedroso[1,2], Mark J F Brown[1*], Seirian Sumner[2,3*]

[1] School of Biological Sciences, Royal Holloway, University of London, Egham, TW20 0EX, UK

[2] Institute of Zoology, Zoological Society of London, Regent's Park, NW1 4RY, UK

[3] School of Biological Sciences, University of Bristol, Woodland Road, Bristol, BS8 1UG, UK

*Equally contributing authors

Corresponding author Inti Pedroso, e-mail: intipedroso@gmail.com

# Abstract

Gene innovation is a key mechanism on the evolution and phenotypic diversity of life forms. There is a need for tools able to study gene innovation across an increasingly large number of genomic sequences to maximally capitalise our understanding of biological systems. Here we present Comparative-Phylostratigraphy, an open-source software suite that enables to time the emergence of new genes across evolutionary time and to correlate patterns of gene emergence with species traits simultaneously across whole genomes from multiple species. Such a comparative strategy is a new powerful tool for starting to dissect the relationship between gene innovation and phenotypic diversity. We describe and showcase our method by analysing recently published ant genomes. This new methodology identified significant bouts of new gene evolution in ant clades, that are associated with shifts in life-history traits. Our method allows easy integration of new genomic data as it becomes available, and thus will be a valuable analytical tool for evolutionary biologists interested in explaining the evolution of diversity of life at the level of the genes.

## 1) Introduction

Understanding the contribution of genome plasticity and gene innovation to phenotypic evolution is a major challenge in evolutionary biology. To date, tests of this fundamental question have been largely restricted to model organisms where high quality genome assemblies were available, for example in species of major interest for biomedical or agricultural research. However, as the costs of DNA sequencing continue to drop and the technologies improve, the repertoire of organisms with sequenced genomes is expanding rapidly to include multiple species of broad taxonomic, ecological and evolutionary interest (Colbourne et al. 2011; Fischman et al. 2011; Gadau et al. 2011). For example, genomes for 10K vertebrate species (Scientists 2009) and 5K arthropods (http://www.arthropodgenomes.org/wiki/i5K) will soon be available. These and other genomic resources will provide unprecedented opportunities to study the relationship between genomic diversity and phenotypic diversity, and potentially change the landscape of the types of questions evolutionary biologists can address. However, in order to maximise the potential of these datasets, a new generation of analytical tools are required to extract meaningful biological patterns from the vast volume of sequence data, stratified across numerous genomes.

The evolution of new genes (i.e., those that lack any significant sequence similarity to other known genes; (Domazet-Loso and Tautz 2003)) plays an important role in generating genomic diversity. It is thought that there are four ways in which new genes can arise, namely gene-duplication, *de novo* origin from non-coding sequences, lateral gene transfer and retrotransposition /(Ding et al. 2012; Zhou and Wang 2008). Gene-duplication and retrotransposition appear to be common mechanisms in eukaryotes and prokaryotes, whilst lateral gene transfer events seem to be very rare in eukaryotes (Zhou and Wang 2008). *De novo* origin was thought to be a rare and unlikely mechanism (Jacob 1977) but recent evidence is changing this view. For example, orphan genes (i.e., genes that appear to be unique to a particular species or clade), are pervasive in every genome sequenced to date (Tautz and Domazet-Loso 2011) and there is increasing evidence that genes often originate *de novo* from non-coding sequences (see review by (Zhou and Wang 2008)) and proto-genes, e.g., in *Saccharomyces cerevisiae* (Carvunis et al. 2012). Overall, the current evidence suggests that *de novo* gene evolution is an important mechanism but its quantitative contribution to gene innovation remains unknown (Ding et al. 2012). The rapid accumulation of high quality genomes provides the raw material to greatly enhance our understanding of when and how genomic diversity is generated.

The evolution of phenotypic innovation appears to coincide with evolutionary periods when new genomic diversity emerges. For example, there is evidence that new gene copies and *de novo* origination of genes accompanied the evolution of embryogenesis and morphology (Domazet-Loso et al. 2007; Khalturin et al. 2008), new cell types (Hemmrich et al. 2012; Arendt 2008) and complex behavioural repertoires (Dai et al. 2008; Emes et al. 2008). Thus, determining whether the timing of gene birth (or family expansion) coincides with the evolution of new phenotypes is a fundamental question in modern biology, as we attempt to understand the general evolutionary processes underlying the diversity of life (Domazet-Loso and Tautz 2008; Domazet-Loso and Tautz 2010a; Hemmrich et al. 2012; Arendt 2008). Current analytical methodologies enable periods of new gene innovation to be detected in a single species (e.g., phylostratigraphy (Domazet-Loso and Tautz 2003)) or gene family expansions in multiple species (e.g., gene-family size analyses (Csuros and

Miklos 2009)). Methods suitable for analysing multiple species do not scale-up well for analyses of tens or hundreds of genomes, and also require re-analysis of the whole dataset as new genomes become available. With the rapid growth in genomic resources, there is increasing demand for new analytical tools that allow comparisons of genomic diversity across a large number of species and evolutionary lineages simultaneously. Such a comparative approach would profit from powerful and well-developed comparative phylogenetic methods (e.g., (Pagel 1999)) to detect macroevolutionary patterns of genomic diversity across multiple lineages and associate them with phenotypic evolution.

In this paper, we develop a comparative method for detecting an important component of genomic diversity – the emergence of new protein families. These new genes are assumed to arise through a model of punctuated evolution via founder genes, which represent evolutionary novelties in protein sequence (i.e. this excludes simple duplications of existing genes or re-shuffling of functional domains) (Domazet-Loso & Tautz 2010). Our method combines an existing approach (phylostratigraphy (Domazet-Loso and Tautz 2003)) with comparative genomics methods to generate the first method and software suite for analysing gene innovation events across whole genomes of multiple species. Thus, it enables systematic mining of gene innovation across the rapidly increasing number of genome sequences available. As proof of concept, we provide a case study test of our method to identify significant leaps in gene innovation that may correspond to the emergence of life-history innovations in the ants. Our software is open source and thus will provide an important step forward for the genomics community in maximal exploitation of omics data sets.

## 2. New Approach: Comparative-Phylostratigraphy

Phylostratigraphy is a statistical method which quantifies gene-founder events across evolutionary periods of a genome by stratifying its genes based on the evolutionary period in which they first appeared (Domazet-Loso and Tautz 2003). A gene-founding event marks the emergence of a protein-coding sequence that forms a new gene lineage or gene family, which may confer functional novelty. Thus, in this sense we define 'new genes' as ones which represent new functional proteins or protein-coding domains, that were not previously present in the genome. This definition assumes the evolutionary origin of a gene through punctuated evolution of protein families via founder gene formation

(Domazet-Loso et al. 2007). Evolutionary periods are referred to as 'phylostrata' and represent points of divergence between a species' lineage and its ancestors. Genes are assigned to a phylostratum when some or all descendants of this ancestor contain the gene, i.e., the gene's phylogeny coalesces into the taxon where the most recent common ancestor (MRCA) of that gene must have existed (Domazet-Loso and Tautz 2003). Phylostratigraphy therefore identifies these new genes and pinpoints their likely point of emergence during specific evolutionary periods (phylostrata). The approach is simply and yet powerful as it helps to overcome some major limitations of current gene-family analyses, such as the need for fully sequenced genomes, the computational burden of building thousands of gene-family trees, resolving orthology versus paralogy relationships, and having to re-build and infer parameters when new genomes are available, whilst still identifying expansion of the gene repertoire (Domazet-Loso et al. 2007; Tautz and Domazet-Loso 2011).

Previously, phylostratigraphy has been applied to study the dynamics of gene innovation through the evolution of a single genome or to compare the relative age of genes expressed in different cells, tissues or developmental stages. These studies have provided major insights into the evolution of human diseases (Domazet-Loso and Tautz 2008; Domazet-Loso and Tautz 2010a), development (Domazet-Loso and Tautz 2010b; Domazet-Loso et al. 2007; Hemmrich et al. 2012) and gene evolution (Domazet-Loso and Tautz 2003; Tautz and Domazet-Loso 2011). But the method has tremendous potential in comparative genomics, to identify general patterns of genomic evolution across species. To realise this potential we need to: a) extend the original methodology to the simultaneous analysis of datasets from multiple species; b) identify the biological functions of the genes experiencing expansions or founding events in different phylostrata; and c) provide an open source software programme for the community to deploy this approach. In this study we achieve these three goals through the development of a new software suite, Comparative-Phylostratigraphy. The remainder of this section explains how this method allows phylostratigraphy to be extended to multiple species (additional details on Supp Material section II) and in Supp Mat section III we present details of the software implementing the original phylostratigraphy and our new methodology.

Comparative-Phylostratigraphy differs from previous methodologies in that it compares simultaneously the dynamics of gene innovation across the evolutionary history of multiple species (Fig 1). The macro-evolutionary comparative component is an important step forward from existing methods (e.g., phylostratigraphy (Domazet-Loso et al. 2007) or ProteinHistorian (Capra et al. 2012)) because it uses phylogenetic information to identify associations between macro-evolutionary trends of gene innovation and species life-history traits. Specifically, it enables the identification of associations between gene innovation in different phylostrata with the emergence of general phenotypic transitions in evolution that are of ecological, economic and evolutionary importance. Such transitions include the shift from free-living to obligate symbioses (e.g., the origin of agriculture in fungus-growing ants), a change in life strategy (e.g., from generalist to specialists in parasitoids) and major evolutionary transition (e.g., the shift from solitary to social living).

Comparative-Phylostratigraphy starts by obtaining phylostratigraphic profiles, i.e., the number of genes assigned to each phylostrata, for each species using phylostratigraphy (additional details on Supp Mat section II-A). The comparative component is achieved by contrasting the phylostratigraphic profiles between species to identify commonalities and differences associated with species' phenotypes. We use a poisson linear mixed model formulated as $c_{fs} \sim N_s * e^{\wedge}(a*1 + \mathbf{b}*\mathbf{t}_s + \mathbf{h}*\mathbf{t}_s*\mathbf{f} + \mathbf{g})$ where $c_{fs}$ is the number of proteins in species $s$ assigned to phylostratum $f$, $N_s$ is the total number of proteins annotated in the genome of species $s$, $\mathbf{t}$ is a vector of species' covariates which can be coded as factors or as continuous variables, $a$ is the model's intercept, $\mathbf{b}$ is a vector of coefficients indicating the effect of species' covariates on the counts $c_{fs}$, $\mathbf{h}$ is a vector of coefficients for the species' covariate and phylostrata interaction, and $\mathbf{g}$ is a vector of species coefficients with variance-covariance matrix $\mathbf{Q} = \tau\mathbf{C}$, with $\mathbf{C}$ defined by the species relationship $\tau$ and the scaling parameter. The $\mathbf{C}$ matrix is calculated from the species tree, using the method of Hadfield and Nakagawa (2010). This model is fitted using MCMCglmm (Hadfield 2010) software for R (R Development Core Team 2012). The coefficients $\mathbf{b}$ can be used to assess a difference between the phylostratigraphic profiles, and the coefficient $\mathbf{h}$ identifies differences associated with specific phylostrata and covariates (additional details on Supp Mat section II-B).

Next, we extended the methodology to identify deviations in the functions of proteins assigned to a particular phylostratum that are different between the species being compared. This corresponds to an expansion of specific biological function at a specific period in evolution. This analysis was implemented by integrating analyses of phylostratigraphic profiles with gene-set enrichment analyses using hypergeometric distribution statistics (see Supp Materials section II-C). Additional details of methods and development are provided in the Supplementary Materials.

To facilitate accessibility and use of our methodology we provide an open source software package 'Comparative-Phylostratigraphy' implementing all these methods (https://github.com/inti/Comparative-Phylostratigraphy). The software includes utility scripts to analyse the data, and produce summary tables and figures to explore the results (additional details on Supp Mat section III).

## 3. Case Study: Gene Innovations with life-history innovations in the ants

### 3.1. Hypotheses

All ants are eusocial (having cooperative brood care, overlapping generations and reproductive division of labour). To date sequenced genomes for seven ant species are publicly available, including two species of fungus-growing or farming ants (Attini, Myrmicinae), *Acromyrmex echinatior* and *Atta cephalotes*. The other five ants are *Camponotus floridanus (*Formicidae*), Harpegnathos saltator (*Ponerinae*), Linepithema humile (*Formicidae*), Pogonomyrmex barbatus (*Myrmicinae*),* and *Solenopsis invicta (*Myrmicinae*),* all of which are non-farming. Thus, the Attines differ significantly from the other 5 ant species in life strategies. The origin of agricultural practices in insects is a major transition in evolution as it represents a shift from a free-living life strategy to an obligate, symbiotic one. Dependency on fungus farming has arisen at least 3 times independently in the insects: in the ants (Hymenoptera, Attini), termites (Blattodea, Termitoidae) and ambrosia beetles (Coleoptera; Curculionoidea). In the Attine ants, it is believed to have evolved at the origin of the lineage, approximately 50 million years ago (Schultz and Brady 2008). The ants cultivate their fungus, providing it with plant material; the fungus breaks down the indigestible plant material and converts it into gongylidia, which the ants feed to their brood. This is an obligate symbiosis, where ants and fungus are mutually dependent on each other. To our knowledge, no Attine ant has abandoned fungus farming to revert to a free-living life-style, testifying to the efficacy of this strategy (Schultz and Brady 2008). To date, we have a very limited understanding of what genome changes contributed to the emergence of agriculture in these ants. Recent genome sequences for two Attines identified gene loss and contractions, and over- and under-representations of specific genes/gene families relative to non-attines (Nygaard et al. 2011; Suen et al. 2011). But, an untested hypothesis is that the origin of agriculture in the Attine ants is associated with a proliferation of *new* genes, rather than simply genomic reductions, which may be expected from their derived, symbiotic lifestyles.

These seven ants also represent other life-history differences, such as different levels of eusociality, dietary specialisations and mating systems (see Supp Table 1 for additional information on species' life-history traits). For example, *P. barbatus* is a seed-harvesting ant that, like the leaf-cutters, exhibits high levels of polyandry, while the monandrous *H. saltator* exhibits only primitive eusociality.

We compare these ant genomes to identify any differences in their phylostratigraphic profiles that may be associated with life-history innovations. Specifically, we ask whether any significant levels of gene innovations in the ants are associated with the shift from non-farming to farming.

## 3.2 Results and Discussion

We used Comparative-Phylostratigraphy (see Methods and Supp Mate sections I and II) to construct and compare the phylostratigraphic profiles of each of the seven ant species. This revealed two main periods of gene innovation. Firstly, most (33-39%) of the ant proteins originated at the origin of cellular life or at the basal branch of eukaryotes and metazoa, as observed for other eukaryotic groups (Tautz and Domazet-Loso 2011) (Fig 2). Most of these genes represent the core cellular machinery, such us central metabolism, DNA replication, RNA-transcription and protein translation, all of which are shared with unicellular organisms. This finding provides proof-of-concept for our method.

A second significant period of innovation, accounting for an additional 8-22% of new proteins, occurred after the origin of the Formicidae (phylostratum) (Fig 2). There was a significant difference between the farming and non-farming ant species in the number of new genes appearing at the phylostrata after the origin of the Formicidae (p-value < 0.008, regression-coefficient = 0.68 (95% confidence intervals: 0.17-1.19)). No other phylostratum differed between the two groups (Supp Table 2), suggesting that the divergence of the farming lineage from the non-farming lineage explains these differences in gene innovation. These results suggest that the genome composition of these two groups of ants differs significantly in the number of genes that originated after the farming and non-farming lineages diverged. We explored this result in more detail by examining for each species the percentage of new genes in each of the phylostrata since the Formicidae ancestor, i.e., the last common ancestor of the two groups. We found a bimodal distribution in the proportion of new genes among species that was explained by the division between non-farming and farming life styles: in non-farming ants 12 ± 4% of genes in the genome are species specific (new) genes, whilst in the farming ants 21 ± 1.5% are new genes (p-value < 0.05).

Finally, we tested whether the genes defining the phylostrata of the two ant life styles differed in their putative biological functions. Five Gene Ontology (GO) categories were identified as differentially enriched between ant life styles, and interestingly, the genes belonging to these GO categories all originated in phylostrata *prior* to the origin of the Formicidae (Table 1). Interestingly, two of the enriched categories are associated with metabolism (GO:0004396 and GO:0008137), suggesting a potential link with the diets associated with the two life strategies.

    These results provide provisional support for the hypothesis that the shift to agriculture in ants was accompanied by the emergence of a substantial number of new genes. This is plausible, given that the functional changes appear to be associated with metabolism, and thus indicate a link to dietary changes. However, we cannot exclude the possibility that gene innovation is additionally (or instead) explained by the shift from single mating by queens to high promiscuity, since *P. barbatus* also showed elevated levels of gene innovation, and queens of *P. barbatus* and the two attines are highly multiply mated, whilst the other species are all obligately singly mated. The evolution of multiple mating is a key innovation in the highly complex eusocial species, conferring multiple life-history benefits (Sumner et al. 2004). Likewise, we cannot exclude the possibility that simply a shift from ancestral scavenging behaviour to reliance solely on plant material can explain the gene innovations. However, our results cannot be explained by level of sociality, colony size or type of coloniality (Supp Table 3). With the addition of genome sequences for the lower attines, which are singly-mated fungus farmers, and further herbivorous species (e.g. *Messor* spp.), our method will allow a fine-level discernment of the relationships between life-history innovations and punctuated gene origination in the ants.

## Summary and Future Directions

These results provide proof of concept for our method and software, through the identification of novel genes at the origin of the Eukaryotes and correlation of gene innovation patterns with species life-history traits. In addition, our results provide the first suggestion that shifts in life-style in ants from an ancestral free-living state to a symbiotic mutualism are associated with a significant increase in gene innovation across the genome. Although interpretation of our case study was limited by the species currently available, our results support similar findings of an association between large-scale gene innovation and major phenotypic innovations (see section 1 above). Interpretation of gene functionality, however, remains limited for our, and similar, studies, as determining the functionality of orphan (new) genes is currently a major challenge. Nonetheless, our methodology provides a step forward by allowing us to assess whether the prevalence of new genes in genomes is associated with a species' trait. This insight could not be gained using previous methodologies, which focus on analysis of a single genome (e.g., phylostratigraphy (Domazet-Loso et al. 2007) or ProteinHistorian (Capra et al. 2012)).

In summary, Comparative-Phylostratigraphy provides an important tool for evolutionary biologists. As more genomes from a wider range of organisms are sequenced, this method will allow finer-scale dissection of life-history traits and identification of the key novel genes that accompany the evolution of major biological innovations. Our methodology scales up to large numbers of genomes without needing to re-compute time-consuming analyses as new genomes become available. Sequence comparisons to public databases are now routine in any genome project, making our method readily applicable as part of mainstream genome annotation/analysis pipelines. Future directions include identifying gene innovations associated with the transition from non-social to social living in the social insects, or the evolution of the parasitic life-style from a non-parasitic one. The methodology can also be applied without species boundaries, for example, to examine gene innovations associated with continuous variables such as body size or brain size. Future methodological challenges include the integration of transcriptome data into analyses, which would greatly extend the range of taxa and questions that can currently be addressed.

## Materials and Methods

Phylostratigraphy and comparative-phylostratigraphy (described in Supp Materials sections II) were performed using our newly developed software Comparative-Phylostratigraphy, which is described in detail in Supp Materials sections III. Species' classification as farming or non-farming ants was obtained from relevant literature (Hölldobler and Wilson 1990). Genome sequences were obtained from the Hymenoptera Genome Database (Munoz-Torres et al. 2011). Species' phylogeny was based on a phylogenetic analysis of protein sequences (see Sup Materials section I).

**Acknowledgements**

This work was supported by The Leverhulme Trust through research grant F/07 537/AK to MJFB and SS. Computing resources were accessed through the RHUL Centre for Systems and Synthetic Biology.

# Figure and Table Legends

**Figure 1. Overview of the Comparative-Phylostratigraphy analyses.** A set of protein sequences are compared with sequence databases and based on the homologous sequences found each protein is assign to a phylostratum. Each grey panel represents one species and analyses within them are performed for all species to generate a phylostratigraphic profile for each protein set. These profiles can be interpreted using within or across species analyses. See main text and Supp Mat section II for details of each step.

**Figure 2: Phylostratigraphic profiles for seven ants.** Phylostratigraphic profiles are presented as the percentage of genes in the genome (y-axis) that can be traced back to each phylostratum (x-axis), with phylostrata being ordered chronologically. Panel A presents data for all species on all phylostrata. Species colour code is shown in the legend. Panel B presents the phylostratigraphic profile from the Formicidae phylostratum forward up to each extant species, with phylostrata being ordered chronologically. Each sub-panel shows data for one specie. We have included the Formicidae phylostratum to aid comparison with Panel A. In all cases confidence intervals were obtained by conducting 5000 bootstrap replicates.

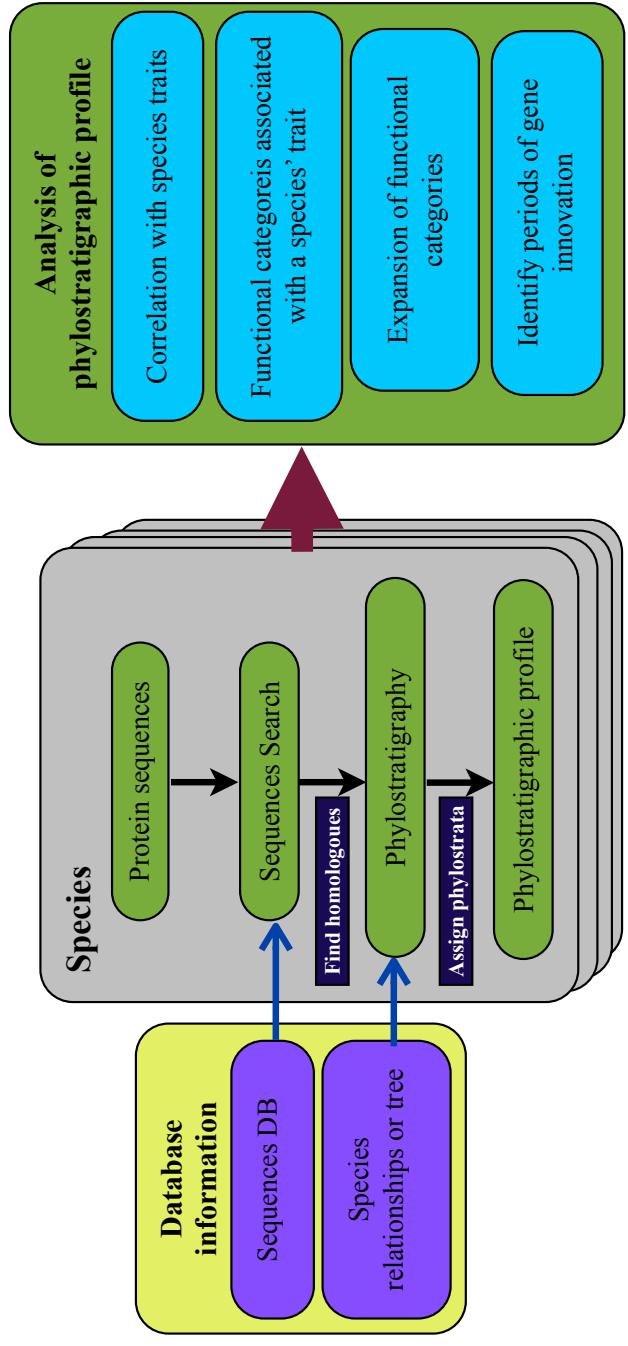

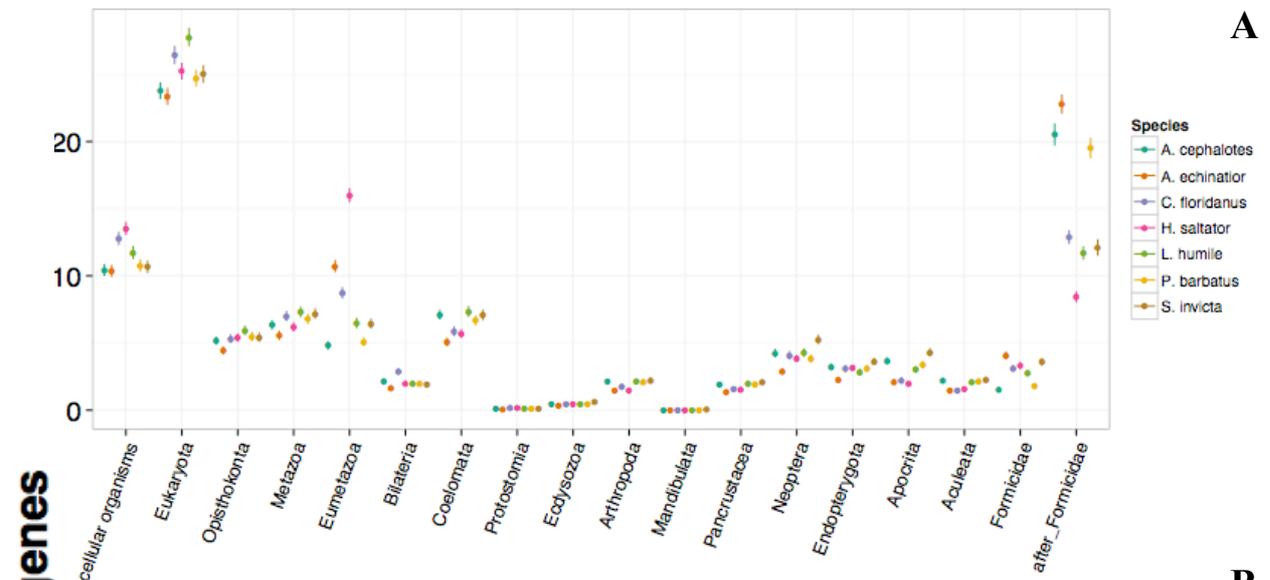

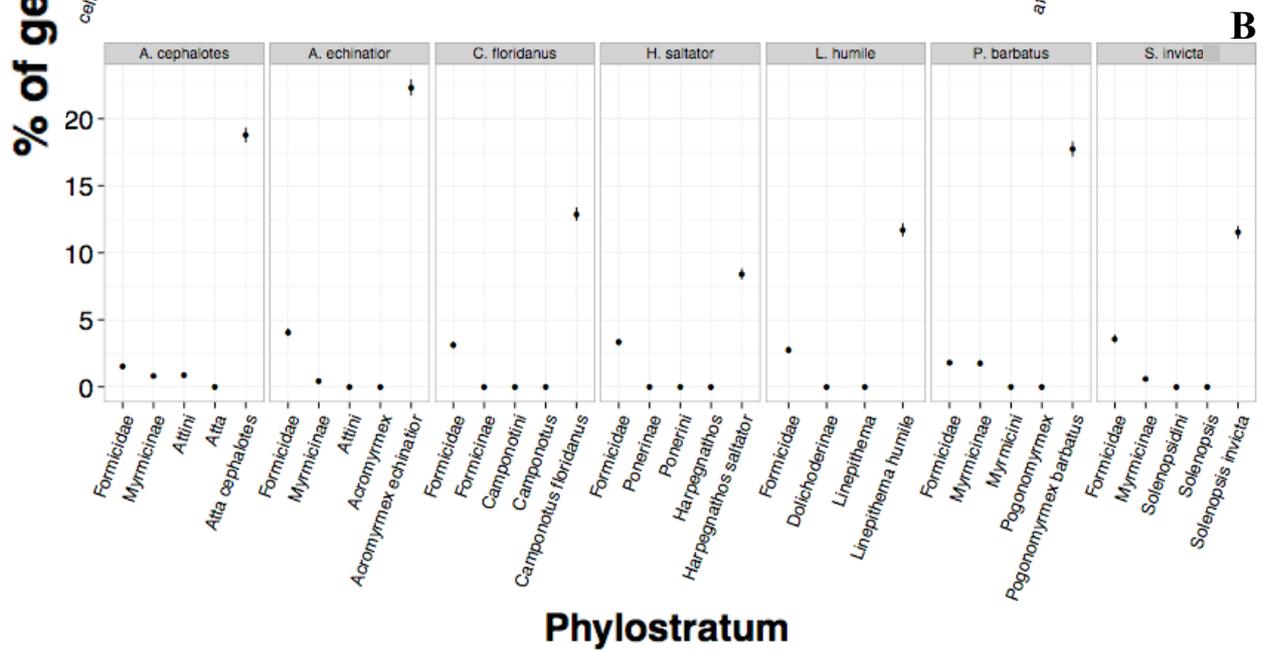

| GO category | Phylostratum | Beta | SE | P-value | local FDR | GO category description |
|---|---|---|---|---|---|---|
| GO:0004396 | Metazoa | -3.05 | 0.15 | 5.7E-06 | 0.010 | Hexokinase activity |
| GO:0051205 | Cellular Organisms | 2.15 | 0.16 | 3.8E-05 | 0.010 | Protein insertion into membrane |
| | Eukaryota | -1.73 | 0.13 | 4.00E-05 | 0.038 | |
| GO:0008137 | Endopterygota | -1.99 | 0.18 | 1.1E-04 | 0.038 | NADH dehydrogenase (ubiquinone) activity |
| GO:0046907 | Opisthokonta | 2.31 | 0.23 | 1.6E-04 | 0.038 | Intracellular transport |

**Table 1. GO categories with local FDR < 0.05 in phylostratum-wide comparisons.** Reported are GO terms with evidence of differential enrichment between leaf-cutter and non-leaf-cutter ants. A positive regression coefficient (Beta) means a higher enrichment in leaf-cutter ants. SE = standard error, P-value = p-values for the hypothesis test of Beta = 0 and local FDR = local False Discovery Rate.

# Supplementary Materials

## Supplementary Methods

### I. Sequences and Sequence Analyses

We obtained predicted protein sequences from the published genome sequences of seven ant species, including two farming ants *Acromyrmex echinatior* and *Atta cephalotes* and five non-farming ants *Camponotus floridanus*, *Harpegnathos saltator*, *Linepithema humile*, *Pogonomyrmex barbatus*, *Solenopsis invicta*, from the Hymenoptera Genome Database (Munoz-Torres et al. 2011). These ants are well-characterised biologically (Supp Table 1).

Protein sequences were annotated to Gene-Ontology (GO) categories (Ashburner et al. 2000) using InterProScan v4.5 (Zdobnov and Apweiler 2001). Protein sequences were compared to the non-redundant protein database from NCBI using BLAST (Altschul et al. 1990). A phylogenetic tree for these species was constructed using protein families with a single predicted protein across all 33 Arthropods species included on the OrthoDB v5 database (Waterhouse et al. 2011), this ensures the families are likely true single-gene families and not an example of a reduction in size in the ancestor of all ants species. The protein sequences were aligned using T-COFFEE (Notredame et al. 2000) by combining the output from multiple software (M-COFFE methodology) with the following options: "-method mafftgins_msa muscle_msa kalign_msa t_coffee_msa". Alignment blocks of at least 10 amino-acids were extracted using GBlocks (Castresana 2000). We utilised ProtTest to identify a model of molecular evolution for the concatenated alignment blocks and the best model selected by both Akaike Information Criteria or Bayesian Information Criteria had substitution rates with a 4-categories gamma distribution, amino-acid matrix = LG and amino-acid frequencies estimated from the data. We used these options to construct a species tree using PhyML (Guindon et al. 2010) with 1000 bootstrap replicates to assess nodes support. The structure of the tree agreed with the known species relationships and no node had low (< 90%) bootstrap support (Supp Figure 1).

### II. Comparative-Phylostratigraphy

a) Phylostratigraphic Analyses

We implemented the phylostratigraphic analyses described by Domazet-Loso et al. (2007) in a newly-developed software tool called Comparative-Phylostratigraphy. We refer the interested reader to Supplementary Files and the software website (https://github.com/inti/Comparative-Phylostratigraphy) for a detailed description of the software implementation and capabilities. We utilised the phylostratigraphy.pl script to parse the BLAST output files from each species analyses separately. This script assigns every protein to a phylostratum (i.e., ancestor) from the query species going back to the last common ancestor of all cellular organisms. Results were filtered by an e-value equal to $1e^{-3}$, which has been shown to provide a good trade off between sensitivity and specificity (Domazet-Loso and Tautz 2003). If a protein did not have BLAST results with e-value $< 1\times10^{-3}$ it was considered to be species specific and assigned to the query species. We removed from further analyses phylostrata without proteins assigned to at least one species. The phylostratigraphy.pl script also provided the total count of proteins mapped to each phylostratum with 95% confidence intervals calculated by bootstrapping with 5000 replicates. For our statistical comparisons (see below) we defined two groups in order to test our main hypothesis - namely the farming and non-farming ants - which have the Formicidae taxon as their last common ancestor; this is oldest the phylostratum at which we can compare them. To allow comparison of gene innovation patterns after their divergence from the Formicidae we combined all genes appearing after the origin of the Formicidae into a single new phylostratum that we called "After_Formicidae".

b) Comparison of phylostratigraphic profiles

For each species we had a phylostratigraphic profile, i.e., the number of proteins the phylostratigraphic analyses mapped to each phylostratum and the total number of proteins in each species. In order to compare these profiles between our two groups of interest we used a Bayesian Poisson linear mixed model formulated as $c_{fs} \sim N_s * e^{\wedge}(a*1 + \mathbf{b}*\mathbf{t}_s + \mathbf{h}*\mathbf{t}_s*\mathbf{f} + \mathbf{g})$ where $c_{fs}$ is the number of proteins of species $s$ assigned to phylostratum $f$, $N_s$ is the total numbers of proteins annotated in the genome of specie $s$, $\mathbf{t}$ is a vector of species' covariates which can be coded as factors or a continuous variables, $a$ is the model's intercept, $\mathbf{b}$ is a vector of coefficients indicating the effect of species' covariates on the counts $c_{fs}$, $\mathbf{h}$ is a vector of coefficients for the species' covariate and phylostrata interaction and $\mathbf{g}$ is a vector of species coefficients with variance-covariance matrix $\mathbf{Q} = \tau\mathbf{C}$ with $\mathbf{C}$ defined by the species relationship and $\tau$ the scaling parameter. The $\mathbf{C}$ matrix is calculated from the species tree, using the method of Hadfield and Nakagawa (2010). This model is fitted using MCMCglmm (Hadfield 2010) software for R (R Development Core Team 2012). The coefficient $\mathbf{b}$ can be used to assess a difference between the phylostratigraphic profiles and the coefficient **h** identifies differences associated with specific phylostratum and covariates. The model was fitted using the MCMCglmm software for R with default parameters with the exception of the offset ($N_s$). MCMCglmm does not explicitly implement the use of an offset but this can be included as a regressor coded as $\log(N_s)$ and setting its prior mean to 1 and variance to $1\times10^{-6}$ to ensure the mean of its coefficient is 1. We ran 1 million iterations with a burn-in of 50000 and sampled parameter values every 1000 iterations after the burn-in to estimate the posterior distribution of the parameters. Model parameters were summarised by their mean and 95% confidence intervals from the 950 parameters' samples from their posterior distribution. Convergence was assessed visually by plotting the value of the parameters over the iterations using the 'coda' software (Plummer et al. 2006) for R (R Development Core Team 2012). Convergence was achieved soon after 100000 iterations with no sign of autocorrelation in the samples.

c) Cross-species gene-ontology comparison of phylostrata

We aimed to identify differences in the function of proteins assigned to a given phylostratum between farming and non-farming species. By combining the GO annotations of the proteomes and the results of the phylostratigraphic analyses we obtained the number of proteins belonging to each GO category in each phylostratum for each species. We filtered out GO categories in a phylostratum-wise manner, i.e., removed them from the analyses of a phylostratum if they did not pass our thresholds but kept them in a phylostratum where they did. The two filters applied were: a) GO categories had to have some variation in their counts in a phylostratum across species; and b) they had to be present in all seven species. After applying these filters, there were 4005 GO categories for statistical analyses. In order to test for differences in GO categories across species at specific phylostrata we follow a two step procedure. Firstly, we calculate the overrepresentation of each GO category in each phylostratum. This is done for each proteome separately. Second, we compare this overrepresentation for each GO category across species using as a null hypothesis that a given GO category has the same enrichment (proportional to its proteome) at a particular phylostratum across farming and non-farming ant species. Thus, we formulated the first step as an enrichment analysis calculated using the hypergeometric distribution, from which we calculated the following z-score:

$$z\text{-}score = \frac{r - n\frac{R}{N}}{\sqrt{n \cdot \left(\frac{R}{n}\right)\left(1-\frac{R}{n}\right)\left(1-\frac{n-1}{N-1}\right)}}$$

where N = total number of proteins in the proteome, R = total number of proteins in a GO category, n = total number of genes in the phylostratum and r = number of genes belonging to the GO category and the phylostratum (i.e., their intersection). We performed cross-species comparisons on the basis of this z-score (i.e., comparing the enrichment directly across species) by regressing the species group (part of the Attini clade or not) and phylogeny on the z-score using a phylogenetic generalised least-square regression as implemented by the 'glm' function of the 'nlme' software for R (R Development Core Team 2012). The coefficient of the farming and non-farming ant covariate can be used to assess the difference in enrichment between species. We used a local False Discovery Rate (FDR, (Strimmer 2008)) of 0.05 to select a set of results for further inspection.

**III. The Comparative-Phylostratigraphy Software Suite**

We developed a complete software suite allowing users to start from results of sequence searches and perform data analyses within and across species (Supp Fig 2). The software suit is composed of a series of command line tools written in Perl , () and R (R Development Core Team 2012). We make use of the PDL, BioPerl and Bio::LITE::Taxonomy Perl libraries (available from CPAN, www.cpan.org), the MCMCglmm (Hadfield 2010), plyr (Wickham 2011) and nlme (Pinheiro J et al. 2012) software packages for R and The National Center for Biotechnology Information (NCBI) E-utilities Access Programatic Interface (Sayers 2009). Before using Comparative-Phylostratigraphy the user needs to perform sequence comparisons between the proteomes of interest and sequence databases. One of several specialised software packages, e.g., BLAST (Altschul et al. 1990) or PARALIGN (Saebo et al. 2005), can be used for this purpose. The Comparative Phylostratigraphy software suite provides scripts to perform individual data analysis tasks. The first step of our pipeline is to parse the sequence search results and calculate the age of every gene. This is done with the phylostratigraphy.pl script that will take as input the NCBI Taxonomy database files, sequence search output files and produce the following output files: a) indicating the query species ancestor in which each gene originated, and b) a summary of the number of genes that originated at each ancestor with confidence intervals calculated by standard bootstrap methods. The results can be used to generate a genome-wide summary of the gene-creation process at different times in the past (see Fig 2 in main text for an example).

Below we describe how our software allows some outstanding problems in the evolution of genomic novelty to be addressed, and how the individual components fit together. We divide the questions broadly into within and across-species questions.

iv) **Within species:** Within species comparisons are those concerned with identifying differential patterns of gene emergence or gene age between sets of genes within the same genome. For instance, one may be interested to know when genes associated with a biological function (e.g., immunity) first appeared; did these genes appear simultaneously at a single point in evolutionary time, or as multiple events over several time periods? Are they younger or older than the genome's average gene? The first and second questions can be answer with the phylostratum_enrichment.pl script, which will calculate the p-value of enrichment of a gene-set of interest among the genes of each phylostratum. Enrichment p-values are calculated using the hypergeometric distribution. This analysis can be performed for one or many gene-sets, facilitating large-scale analyses, e.g., using Gene Ontology (Ashburner et al. 2000) categories. The third question can be addressed with the gene-set age index (GAI) = $(\sum ps_i * w_i)/\sum w_i$, where the sums are calculated over the total number of genes in the gene-set, $ps_i$ is the index of the phylostratum index of the $i$th gene (increasing from 0 for "cellular organism" to the extant taxa being analysed) and $w_i$ is a gene specific weight. The transcriptome age index (TAI) is a special version of GAI in which each gene's weight is represented by its expression level (Domazet-Loso and Tautz 2010). Both GAI, TAI and statistical comparisons using them can be computed with the gene_set_age_index.pl.

v) **Cross-species:** The approach can be expanded to multi-species comparisons, to determine whether biological innovations, shared by groups of species, are associated with a synchronised emergence of specific set(s) of new genes. In order to test for differences in the number of genes originating in a particular phylostratum across two or more species we use a generalised linear mixed model accounting for species non-independence (using the species phylogeny) and the different number of genes in its genome. In addition, if a gene's annotation is available, it is possible to test whether the differences of gene content between species are due to genes belonging to a pre-defined gene-set originating in a particular phylostratum. The combination of these two analyses has the potential to identify the evolutionary origin and function of genes associated with biological differences, not just at the species level, but among groups of species sharing a specific phenotypic trait. One would normally be interested in grouping species by some characteristic, although this is not a requirement, and testing differences between these groups at each phylostratum.

# Supplementary Figures and Tables

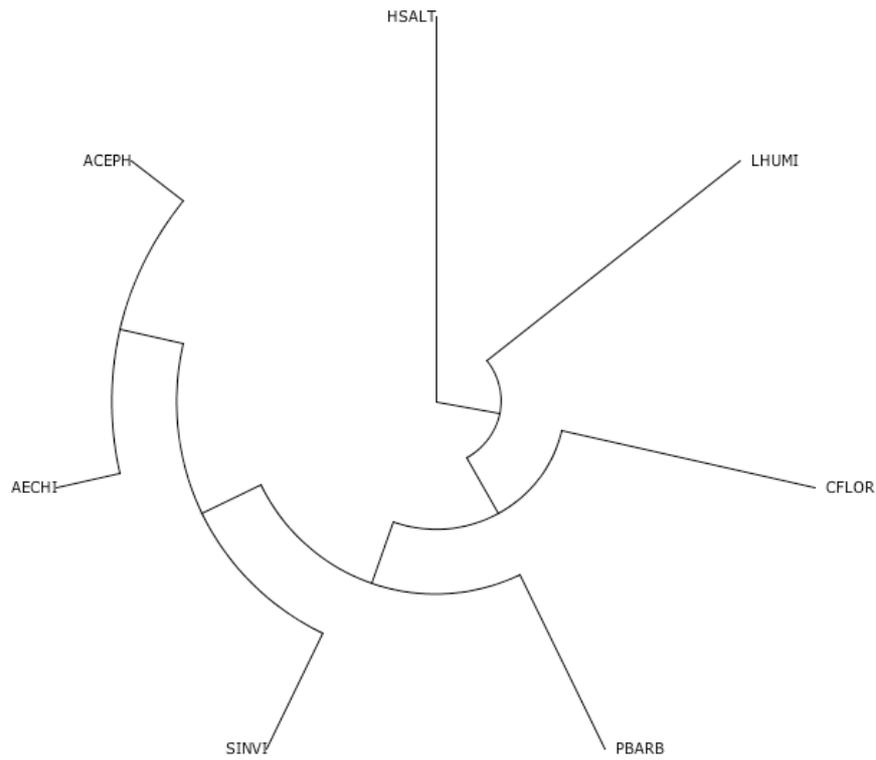

**Supplementary Figure 1. Phylogenetic tree of ant species used on this study.** Aech= *Acromyrmex echinatior*; Aceph= *Atta cephalotes*; SINVI=*Solenopsis invicta*; PBARB= *Pogonomyrmex barbatus*; CFLOR=*Camponotus floridanus*; LHUMI=*Linepithema humile;* HSALT= *Harpegnathos saltator*

| Species | Diet | Level of eusociality | Worker polymorphism | Colony sizes | Colony type | Mating |
|---|---|---|---|---|---|---|
| *A. echinatior* | fungus-growers | complex | polymorphic | 1000s to 1000000s | Multi-colonial | highly polyandrous |
| *A. cephalotes* | fungus-growers | complex | polymorphic | 1000s | Multi-colonial | highly polyandrous |
| *C. floridanus* | omnivores | complex | polymorphic | 1000s | Multi-colonial | monandrous |
| *P. barbatus* | seed-harvesters | complex | monomorphic | 1000s | Multi-colonial | highly polyandrous |
| *L. humile* | omnivores | complex | monomorphic | 1000000s | Unicolonial | monandrous |
| *S. invicta* | omnivores | complex | polymorphic | 100000s | Unicolonial/ Multi-colonial | monandrous |
| *H. saltator* | carnivores | simple | monomorphic | 100s | Multi-colonial | monandrous |

**Supplementary Table 1. Life history traits of ant species used in this study.**

| Phylostratum | Mean | l-95% | u-95% | pMCMC |
|---|---|---|---|---|
| Eukaryota | 0.046033 | -0.428018 | 0.605385 | 0.85895 |
| Opisthokonta | 0.002264 | -0.541311 | 0.493449 | 0.98947 |
| Metazoa | -0.013792 | -0.571301 | 0.460586 | 0.97474 |
| Eumetazoa | 0.050555 | -0.459542 | 0.572301 | 0.84421 |
| Bilateria | 0.008926 | -0.557228 | 0.554180 | 0.96632 |
| Coelomata | 0.055190 | -0.430008 | 0.615158 | 0.84211 |
| Protostomia | -0.165075 | -0.741127 | 0.562184 | 0.60211 |
| Ecdysozoa | -0.054907 | -0.648337 | 0.435153 | 0.87158 |
| Arthropoda | 0.049829 | -0.471604 | 0.570412 | 0.85474 |
| Mandibulata | -0.129663 | -1.064911 | 0.837158 | 0.80632 |
| Pancrustacea | 0.011370 | -0.559883 | 0.508682 | 0.97684 |
| Neoptera | -0.059579 | -0.586045 | 0.473948 | 0.78316 |
| Endopterygota | -0.018588 | -0.499971 | 0.540163 | 0.97474 |
| Apocrita | 0.114343 | -0.442798 | 0.621413 | 0.66947 |
| Aculeata | 0.093462 | -0.434946 | 0.602015 | 0.72000 |
| Formicidae | 0.016648 | -0.524736 | 0.476239 | 0.92421 |
| After_Formicidae | 0.687301 | 0.167685 | 1.185113 | 0.00842 |

**Supplementary Table 2. Comparison of phylostratigraphic profiles in farming versus non-fungus-growing ants.** pMCMC represent the number of times the mean value of coefficient was below or above 0 (for positive and negative coefficients, respectively) across the 950 samples from the posterior. Phylostrata are ordered chronologically.

| Specie trait | Regression coefficient | | | p-value |
|---|---|---|---|---|
| | Mean | lower CI | upper CI | |
| Farming/non-faming | 0.687301 | 0.167685 | 1.185113 | 0.00842 |
| polyandrous/monandrous | 0.780652 | 0.341857 | 1.253769 | < 0.001 |
| Eusociality level | -0.845302 | -1.430716 | -0.210889 | 0.00632 |
| Colony size | 0.102858 | -0.057373 | 0.251377 | 0.189 |
| Colony type | 0.14155 | -0.18993 | 0.48217 | 0.465 |

**Supplementary Table 3. Association between species traits and the gene-innovation on the After-Formicidae phylostratum.** pMCMC represent the number of times the mean value of coefficient was below 0 across the 950 samples from the posterior. Comparison were made using as reference Farming for farming/non-farming, simple for eusociality level and polyandrous for polyandrous/monandrous. Colony type was coded as unicolonial = -1, unicolonial/multi-colonial = 0, multi-colonial = 1 and colony size as 100s = 2, 1000s = 3, 100000s = 4, 1000000s = 6. Please that the result of 'Eusociality level' suggest a reduction in gene-innovation for *H. saltator*, the only simple eusocial specie, which does not explain out patter of interest, i.e., increase gene-innovation on farming ants.